# Spin Hall torque magnetometry of Dzyaloshinskii domain walls

Satoru Emori[1], Eduardo Martinez[2], Kyung-Jin Lee[3,4], Hyun-Woo Lee[5],

Uwe Bauer[1], Sung-Min Ahn[1], Parnika Agrawal[1], David C. Bono[1], and Geoffrey S. D. Beach[1*]

[1] Department of Materials Science and Engineering, Massachusetts Institute of Technology,

Cambridge, Massachusetts 02139, USA

[2] Dpto. Física Aplicada. Universidad de Salamanca,

Plaza de los Caidos s/n E-38008, Salamanca, Spain

[3] Department of Materials Science and Engineering, Korea University, Seoul 136-701, Korea

[4] KU-KIST Graduate School of Converging Science and Technology, Korea University, Seoul

136-713, Korea

[5] PCTP and Department of Physics, Pohang University of Science and Technology, Kyungbuk

790-784, Korea

**Current-induced domain wall motion in the presence of the Dzyaloshinskii-Moriya interaction (DMI) is experimentally and theoretically investigated in heavy-metal/ferromagnet bilayers. The angular dependence of the current-induced torque and the magnetization structure of Dzyaloshinskii domain walls are described and quantified simultaneously in the presence of in-plane fields. We show that the DMI strength depends strongly on the heavy metal, varying by a factor of 20 between Ta and Pa, and that strong DMI leads to wall distortions not seen in conventional materials. These findings provide essential insights for understanding and exploiting chiral magnetism for emerging spintronics applications.**

*email: gbeach@mit.edu

**I. Introduction**

Spin-orbit-driven phenomena at heavy-metal/ferromagnet (HM/FM) interfaces have become the focus of intense research efforts. The influence of spin-orbit coupling (SOC) on spin transport and magnetization textures leads to new fundamental behaviors that can be exploited in high-performance, low-power spintronic devices.[1–23] In HM/FM bilayers, Rashba[3,5,6] and spin-Hall effects[7–9,14–16] can generate current-induced spin-orbit torques (SOTs)[24] potentially much stronger than conventional spin-transfer torques (STTs).[25] In these same materials, SOC and broken inversion symmetry[15,16,26–30] can stabilize chiral spin textures such as spin spirals,[26] skyrmions,[27,30] and homochiral DWs[15,16,28,29] through the Dzyaloshinskii-Moriya interaction (DMI).[29–32] The influence of SOTs on chiral spin textures has only begun to be explored, but recent work suggests spin torque from the spin Hall effect (SHE) can drive DMI-stabilized homochiral Néel DWs with very high efficiency.[15,16] The behavior of these Dzyaloshinskii DWs[31] is however not yet well-understood, due in part to the difficulty of disentangling spin torques and spin textures in these materials.

Here we exploit the angular dependence of the SHE torque to quantify its role in DW dynamics while simultaneously probing the structure and energetics of Dzyaloshinskii DWs. We find that the DMI in HM/CoFe bilayers depends strongly on the HM, but its dependence is distinct from that of the SHE. The DMI exchange constant differs by a factor of ~20 between Ta and Pt, but has the same sign, whereas the SHE for these HMs is of similar magnitude but opposite sign. These results show that while the SHE and DMI both derive from spin-orbit coupling, they arise from distinct mechanisms in these materials and can be independently engineered. Moreover, we uncover a qualitatively new behavior exhibited by DWs in the presence of strong DMI, wherein torque applied to a DW rotates not just the DW moment, but tilts the entire DW line profile. We describe this unconventional behavior through analytical and micromagnetic modeling that accurately describes our experiments and permits quantitative extraction of the DMI strength in such materials. These results provide fundamental insight into interface-driven chiral magnetism and guidance for designing SOC-enabled spintronic devices.

## II. Experiments

We studied perpendicularly-magnetized Pt/CoFe/MgO and Ta/CoFe/MgO nanostrips that served as conduits for DWs. Ta(3nm)/Pt(3nm)/CoFe(0.6nm)/MgO(1.8nm)/Ta(2nm) and Ta(5nm)/CoFe(0.6nm)/MgO(1.8nm)/Ta(2nm) films were sputter-deposited onto $Si/SiO_2$ substrates at room temperature, as described in Ref. 15. Vibrating sample magnetometry on continuous films revealed full out-of-plane remnant magnetization and in-plane (hard-axis) saturation fields $H_\perp \approx 3$ kOe for Ta/CoFe/MgO and $H_\perp \approx 5$ kOe for Pt/CoFe/MgO. The saturation magnetization $M_s$ was $\approx 700$ emu/cm$^3$ for both films.

The films were patterned using electron beam lithography and lift-off to produce nanostrips with Ta/Cu electrodes (Fig. 1a) in two separate lithographic steps. Using this device structure, DWs were nucleated by the Oersted field from a 30-ns long 50 mA current pulse injected through the nucleation line (connected to PG1 in Fig. 1(a)), and driven along the nanostrip by a combination of out-of-plane field $H_z$ and electron current density[33] $j_e$ (output by PG2 in Fig. 1(a)). $H_z$ was generated by an air-coil, whereas in-plane bias fields were generated by an iron-core electromagnet.[34] Magnetization reversal was detected locally with the polar magneto-optical Kerr effect (MOKE) using a ~3 μm laser spot. Most measured strips were 500 nm wide; some measurements were conducted on 1200-nm wide strips with identical results (see Supplemental Material).

Fig. 1b shows magnetization switching probed near the center of a nanostrip as $H_z$ was swept at 17 Hz in a triangular waveform. These measurements were obtained through signal averaging of 100 reversal cycles. The dotted line corresponds to a simple hysteresis loop in the absence of nucleation pulses, such that the switching field corresponds to the threshold for random nucleation. The solid line in Fig. 1b shows a similar measurement obtained when short nucleation pulses were applied at the zero-crossings of the swept field $H_z$. In this case, the switching field decreased significantly, and corresponds to the propagation field $H_{prop}$ required to drive the nucleated DW through the defect potential landscape to the probe laser spot.

Fig. 2 shows that $H_{prop}$ varies linearly with $j_e$, indicating current acts as an easy-axis effective field $\vec{H}_{eff} = \chi j_e \hat{z}$ that can assist or hinder DW propagation. Both nonadiabatic STT (Refs. 2,35,36) and spin torque from the SHE (Refs. 19–21,37) generate effective fields of this form, but differ in the dependence of $\chi$ on the DW configuration. For one-dimensional (1D) DWs with the usual Walker profile,[31] $\chi_{SHE} = \frac{\pi}{2} \chi_{SHE}^0 \cos(\Phi)$ for SHE-torque, where $\Phi$ is the angle between the DW moment and the $x$-axis. Here, $\chi_{SHE}^0 = \hbar \theta_{SH} / 2\mu_0 |e| M_s t$, where $\theta_{SH}$, $e$, $M_s$, and $t$ are the spin Hall angle, electron charge, saturation magnetization, and ferromagnet thickness, respectively. By contrast, nonadiabatic STT (Ref. 25) is independent of $\Phi$, with $\chi_{STT} = \pm \hbar \beta P / 2\mu_0 |e| \Delta M_s$, where $\Delta$ is the DW width, $\beta$ is the nonadiabicity parameter, and positive (negative) corresponds to up-down (down-up) DWs such that current drives them in the same direction.

The relative contributions of STT and SHE-torque to $\chi$ can be determined by applying in-plane fields, which by themselves do not move DWs but can reorient $\Phi$. Fig. 2 shows $H_{prop}$ versus $j_e$ for Ta/CoFe/MgO without and with bias fields along $\hat{x}$ and $\hat{y}$. With no in-plane field (Fig. 2d), $j_e$ assists DW motion along the electron flow direction, identically for up-down and down-up DWs. Under large $H_x<0$ (Fig. 2e), $\chi$ changes sign for down-up DWs, while for up-down DWs $\chi$ is unchanged. For large $H_x>0$, the opposite behavior is observed, while for both DW types, $\chi$ tends to zero under large $|H_y|$ (Fig. 2f).

The sign reversal of $\chi$ under $H_x$, and its vanishing under $H_y$, show that the symmetry of $\vec{H}_{eff}$ is consistent with the Slonczewski-like damping-like torque from the SHE. Under large $|H_y|$ Bloch DWs ($\cos(\Phi)=0$) are preferred (Fig. 2c), and the contribution to $\chi$ from the SHE should vanish (Fig. 2f). In this case only $\chi_{STT}$ remains, which according to Fig. 2f is negligibly small. Under large $H_x$ Néel DWs ($\cos(\Phi)=\pm1$) are stabilized with opposite chirality for up-down and down-up transitions (Fig. 2b). In this case $\vec{H}_{eff}$ from the SHE should drive these DWs in opposite directions, as observed experimentally in Fig. 2e. Therefore, since at $H_x=H_y=0$ the SHE-torque drives up-down and down-up DWs identically in the same direction (Fig. 2d), they must be spontaneously Néel with oppositely-oriented internal moments arranged in a left-handed chiral texture (Fig. 2a), consistent with an internal chiral effective field arising from the DMI.[15,28]

### III. Results

### III.a. Weak DMI Case – Ta/CoFe/MgO

Figs. 3a,b show the full field dependence of $\chi$ in Ta/CoFe/MgO. These data were fitted to the form $\chi_{SHE} = \frac{\pi}{2} \chi^0_{SHE} \cos(\Phi)$, where the dependence of the angle $\Phi$ between the DW moment and the x-axis was computed as a function of $H_x$ and $H_y$ using a 1D DW model with

DMI. In this model,[31] the DW surface energy density σ in the presence of in-plane fields can be written

$$\frac{\sigma}{2\Delta\mu_0 M_s} = \frac{1}{2}H_k \cos^2(\Phi) - \frac{\pi}{2}H_D \cos(\Phi) - \frac{\pi}{2}H_x \cos(\Phi) - \frac{\pi}{2}H_y \sin(\Phi) + H_\perp, \quad (1)$$

where $H_k$, $H_D$, and $H_\perp$ are the DW shape anisotropy field, the DMI effective field, and the perpendicular anisotropy field, respectively. The shape anisotropy term accounts for the DW demagnetizing energy, and has an easy-axis along $\hat{y}$ that prefers Bloch DWs.[38] The DMI effective field takes the form $H_D = \pm D/\mu_0 M_s \Delta$ directed normal to the DW, where $D$ is the effective DMI constant and + (-) corresponds to up-down (down-up) DWs. This term prefers homochiral Néel DWs.

By minimizing σ with respect to Φ, one obtains analytical expressions for the dependence of $\cos(\Phi)$, and hence $\chi_{SHE}$, on $H_x$, $H_y$. Since $\cos(\Phi)$ is simply the x-component $m_x$ of the normalized DW internal moment, the SHE can thus be used to probe the DW configuration under in-plane applied fields. In the case of $H_x$ we find

$$\cos(\Phi) = m_x = \begin{cases} +1 & (H_D + H_x) > \frac{2}{\pi}H_k \\ \pi(H_D + H_x)/2H_k & -\frac{2}{\pi}H_k < (H_D + H_x) < \frac{2}{\pi}H_k \\ -1 & (H_D + H_x) < -\frac{2}{\pi}H_k \end{cases}. \quad (2)$$

The dependence of $\cos(\Phi) = m_x$ on $H_y$ can likewise be found through the relation

$$-H_k m_x \sqrt{1-m_x^2} + \frac{\pi}{2}H_D \sqrt{1-m_x^2} - \frac{\pi}{2}H_y m_x = 0. \quad (3)$$

The solid lines in Figs. 3a,b show fits of the data to $\chi_{SHE} = \frac{\pi}{2}\chi_{SHE}^0 \cos(\Phi)$, with $\cos(\Phi)$ determined through Eqs. (2),(3). This simple model accounts quantitatively for the experimental results, yielding best-fit parameters $H_k = 110$ Oe, $|H_D| = 80$ Oe, and $\frac{\pi}{2}\chi^0_{SHE}=15$ Oe/$10^{10}$A/m$^2$, and a chirality corresponding to left-handed Néel DWs.

A remarkable aspect about Fig. 3a is that the spin torques and DW energy terms can be directly and independently read from the figure even without recourse to fitting. The curve for each DW is analogous to a biased hard-axis hysteresis loop, where the horizontal breadth of the transition gives the DW shape anisotropy field $H_k$, and the zero-crossing field gives $H_D$ (see labels in Fig. 3a). Likewise, the spin torques can be read directly from the vertical axis. The amplitude of the measured $\chi$ versus $H_x$ curve is proportional to $\theta_{SH}$, and the vertical offset is proportional to $\beta P$. The symmetry of the SHE effective field can thus be used to probe the orientation of the DW moment $\hat{m}$ under $H_x$ or $H_y$, from which the angular-dependent DW energy terms can be extracted in analogy with conventional magnetometry.

The measured $\chi^0_{SHE}$ corresponds to an effective spin Hall angle $\theta_{SH} \approx -0.11$, in agreement with Ref. 8. Any contribution to $\chi$ by $\chi_{STT}$ would give rise to a field-independent vertical offset in Fig. 3a. We find no such offset within the experimental uncertainty, which implies an upper limit $\beta P = 0.02 \pm 0.02$. This is in contrast to $\beta P > 1$ inferred in similar materials where variations under in-plane fields were not considered.[2,35,36] These results for the first time disentangle SHE and STT in such materials, and show that here nonadiabatic STT plays a negligible role.

In addition to the effective fields $H_k$, $H_D$, the data in Figs. 3a,b can be combined with conventional magnetometry to yield both the DMI constant $D$ and the ferromagnetic exchange constant $A$. The shape anisotropy field for Néel DWs[39] is $H_k \approx M_s t \ln(2)/\pi\Delta$. Here, $\Delta = \sqrt{A/K_{u,eff}}$ is the DW width, and $K_{u,eff}$ is the effective perpendicular anisotropy energy density. Using $M_s$ and $K_{u,\text{eff}}$ determined by vibrating sample magnetometry, we find $\Delta \approx 10.6$ nm, which yields $A \approx 1.0 \times 10^{-11}$ J/m and $D \approx -0.053$ mJ/m$^2$, where the sign indicates left-handedness. These simple measurements hence provide quantitative insight into every relevant micromagnetic parameter simultaneously, which has never before been achieved.

**III.b. Strong DMI Case – Pt/CoFe/MgO**

Figs. 3c,d show $\chi$ versus in-plane fields for Pt/CoFe/MgO, extracted from the slope of $H_{prop}$ versus $j_e$ as was done for Ta/CoFe/MgO. The behavior is qualitatively similar to that in Ta/CoFe/MgO, with $\chi$ changing sign under large $H_x$, and decreasing smoothly with increasing $|H_y|$. However, much larger fields $|H_x| \approx 2000$ Oe are required to reverse $\chi$ (Fig. 3c), and under $H_y$ the decline in $\chi$ is quite gradual (Fig. 3d), suggesting that here the DMI is much stronger. As described below, in the case of strong DMI, the usual rigid 1D DW model cannot adequately describe the response of DWs to torques. First, experimentally probing strong Dzyaloshinskii DWs requires application of large in-plane fields, which not only rotate the DW moment but also cant the magnetization in the domains. Second, since strong DMI pins the DW moment to the DW normal, a torque on the DW moment tends to rotate the DW normal in the film plane, causing the DW to tilt with respect to the nanostrip axis. Below, we first treat these effects separately using an analytical treatment that provides physical insight, and then use a general micromagnetic approach that accounts for both effects numerically in order to quantitatively fit the data in Fig3c,d and accurately extract the DMI strength.

### III.c. Domain Canting and Thiele Effective Forces

When $H_x$, $H_y$ are comparable to the perpendicular anisotropy field $H_\perp \approx 5000$ Oe, the domains cant significantly from $\pm \hat{z}$ and are no longer collinear. Since the magnetization rotation across the DW is then different from 180°, the Walker ansatz no longer applies, and the 1D equations of motion in Ref. 31 must be amended.

The domain wall (DW) profile in the presence of an in-plane field was derived in Refs. 40–43, from which we obtain the Thiele's equations (see Appendix A) that describe the DW dynamics in terms of position $q$ and wall angle $\Phi$. The Thiele force equation under longitudinal field $H_x$ is given by

$$\frac{\alpha}{\Delta}\left(1 - \frac{2h\xi}{\sqrt{1-h^2}}\right)\dot{q} \mp \sqrt{1-h^2}\,\dot{\Phi} = \mp\gamma\sqrt{1-h^2}\,H_z \mp 2\gamma\xi\,H_{SHE}\cos\Phi, \tag{4}$$

and the Thiele torque equation is given by

$$\mp\frac{\sqrt{1-h^2}}{\Delta}\dot{q} - \alpha\sqrt{1-h^2}\left(\sqrt{1-h^2} + 2h\xi\right)\dot{\Phi} = $$
$$2\gamma\sqrt{1-h^2}\,\xi H_x \sin\Phi \mp \gamma\frac{2D}{\Delta\mu_0 M_S}\xi\sin\Phi - \sqrt{1-h^2}\,\gamma\frac{K_s}{\mu_0 M_S}\left(\sqrt{1-h^2} + 2h\xi\right)\sin 2\Phi \tag{5}$$

Here, the upper (lower) sign corresponds to the up-down (down-up) DWs, $H_{SHE} = \frac{\pi}{2}\chi^0_{SHE}j_e$ and $\xi = \tan^{-1}((1-h)/\sqrt{1-h^2})$, with $h \equiv H_x/H_\perp$.

In the case of a transverse field $H_y$, we define $h \equiv H_y/H_\perp$ and find that the Thiele force equation is unchanged from (4), while the torque equation becomes

$$\mp\frac{\sqrt{1-h^2}}{\Delta}\dot{q} - \alpha\sqrt{1-h^2}\left(\sqrt{1-h^2} + 2h\xi\right)\dot{\Phi} = $$
$$-2\gamma\sqrt{1-h^2}\,\xi H_y \cos\Phi \mp \gamma\frac{2D}{\Delta\mu_0 M_S}\xi\sin\Phi - \sqrt{1-h^2}\,\gamma\frac{K_s}{\mu_0 M_S}\left(\sqrt{1-h^2} + 2h\xi\right)\sin 2\Phi \tag{6}$$

By setting $h=0$ in Eqs. (5) and (6), one recovers the conventional Thiele equations without domain canting given in Ref. 31.

One sees from Eq. (4) that the spin Hall effective field is equivalent to an easy-axis applied field given by

$$H_z^{eff} = \frac{2\xi}{\sqrt{1-h^2}} H_{SHE} \cos\Phi \tag{7}$$

so that

$$\chi_{SHE} = \frac{2\xi}{\sqrt{1-h^2}} \chi_{SHE}^0 \cos\Phi \tag{8}$$

With no in-plane field, Eq. (8) reduces to $\chi_{SHE} = \frac{\pi}{2} \chi_{SHE}^0 \cos(\Phi)$ as above, but domain canting due to applied in-plane fields modifies current-induced effective field from the SHE.

The dotted curves in Fig. 3c show $\chi_{SHE}$ versus $H_x$ computed in this model, where the dependence of $\cos(\Phi)$ on $H_x$ can be expressed analytically (see Appendix A). We used the measured $H_\perp$ and set $H_k$=150 Oe based on Taresenko's expression,[39] leaving $H_D$ as the only free parameter. Taking $|H_D|=1800$ Oe for left-handed DWs, this model reproduces the zero-crossing field of $\chi$ versus $H_x$. Interestingly, domain canting leads to the counterintuitive result that in the field range where the DW remains Néel, $\chi$ is diminished when $H_x$ is parallel to the DW moment, and is enhanced when $H_x$ opposes the internal DW moment. Hence, one should expect a quasilinear variation of $\chi$ about $H_x$=0, the slope of which yields the DW chirality.

While this model explains the observed reduction of $\chi$ when $H_x$ is parallel to $H_D$, it predicts a relatively abrupt reversal of DW chirality when $H_x \approx -H_D$, whereas experimentally, $\chi$ changes sign much more gradually. This behavior arises from a tilting of the DW line profile in the plane due to strong DMI, which we treat in the following section.

**III.d. Domain Wall Tilting Under Strong DMI**

The micromagnetic simulations in Fig. 4 reveal the source of the discrepancy between the data in Figs. 3c,d and the rigid 1D model. These simulations were performed using custom code[44] modified to include the DMI (Appendix B) with $D$ = -1.2 mJ/m$^2$, as determined in the next section. The computed sample was 500 nm wide and 0.6 nm thick, and its length was 2048 nm with appropriate boundary conditions imposed to simulate an infinitely long strip. The material parameters used were: exchange constant $A$ = 10$^{-11}$ J/m; saturation magnetization $M_s$ = 7×10$^5$ A/m; and perpendicular magnetocrystalline anisotropy constant $K_u$ = 4.8×10$^5$ J/m$^3$. These parameters corresponded to experimentally determined values for Pt/CoFe/MgO, except for $A$, which was determined experimentally for Ta/CoFe/MgO in Sec. IIIa and is assumed to be the same for Pt/CoFe/MgO. This value of $A$ gives a DW width $\Delta = \sqrt{A/K_{u,eff}}$ =7.6 nm, where $K_{u,eff} = K_u - \tfrac{1}{2}\mu_0 M_s^2$.

At $H_x$=$H_y$=0, the DW spans the nanostrip orthogonally to minimize elastic line energy (Fig.4a), but under in-plane applied fields that tend to rotate the DW moment, the DW line tilts dramatically in the *x-y* plane (Figs. 4b,c). This remarkable behavior can be understood from simple energy minimization under strong DMI. The Zeeman energy tends to align the DW moment $\hat{m}$ with the applied field, while the DMI prefers $\hat{m}$ to remain normal to the DW. If the DW line were fixed rigidly in position, then $\hat{m}$ would rotate progressively towards the applied field at the expense of the DMI energy. However, if the DW line itself rotates in the *x-y* plane, $\hat{m}$ can follow the applied field while dragging with it the DW normal, thereby reducing the DMI energy penalty. Despite the energy cost of increasing the DW length, DW tilting should lower the net energy if the DMI is sufficiently strong.

Indeed, unexplained tilting of current-driven DWs was recently observed[11] in Pt/Co/Ni/Co/TaN. This behavior is fully consistent with strong DMI, which should lead to DW tilting whenever a torque tends to rotate the DW moment in the plane. Under the large currents (~$10^{12}$ A/m$^2$) used in Ref. 11, the SHE effective field exerts a torque on the DW moment about the z-axis which, due to strong DMI at the Pt/Co interface,[16] should cause dynamical tilting of the DW normal, consistent with Ref. 11. In the present case, we apply much smaller currents (~$10^{10}$ A/m$^2$) and examine quasistatic DW motion (thermally-activated propagation through fine-scale disorder), so that the propagating DW configuration under $H_x$, $H_y$ is determined by total energy minimization.

We modeled DW tilting analytically by parameterizing the DW by two angles (Fig. 4d): $\eta$, the angle between $\hat{m}$ and the DW normal, and $\theta$, the tilting of the DW normal away from the x-axis. We assume the DW remains straight, so $\hat{m}$ is everywhere inclined by $\Phi = \eta + \theta$ from the x-axis, and domain canting is neglected for simplicity. The DW energy $E$ under $H_x$ and $H_y$ is then modified from the form in Eq. (1) as

$$E \propto \frac{1}{\cos\theta}\left[\frac{1}{2}H_k \cos^2(\eta) - \frac{\pi}{2}H_D \cos(\eta) - \frac{\pi}{2}H_x \cos(\eta+\theta) - \frac{\pi}{2}H_y \sin(\eta+\theta) + H_\perp\right] \quad (9)$$

which yields the quasistatic DW configuration through minimization with respect to $\eta$ and $\theta$. The out-of-plane anisotropy field is given by $H_\perp = 2K_{u,eff}/\mu_0 M_s$ and accounts for the DW internal energy. In the case of strong DMI, $H_k$ can be neglected when $H_k \ll |H_D|$.

Minimizing Eq. (9) with respect to $\theta$ and $\eta$ yields the equilibrium DW configuration under $H_x$ and $H_y$. In the case of $H_x$, one finds

$$\sin\theta = \frac{\frac{\pi}{2}H_x \sin\eta}{-\frac{1}{2}H_k \cos^2\eta + \frac{\pi}{2}H_D \cos\eta - H_\perp} \quad (10)$$

and

$$\sin\eta = \frac{\frac{\pi}{2}H_x \sin(\eta+\theta)}{H_k \cos\eta - \frac{\pi}{2}H_D}. \tag{11}$$

For $H_x$ far from $H_x = -H_D$, the physical solutions are $\eta = 0, \pi$, and $\theta = 0$. This corresponds to the DW moment orienting along the $x$-axis ($m_x = \pm 1$) with no tilting of the DW normal.

Under transverse field $H_y$, minimizing Eq. (9) with respect to $\eta$ and $\theta$ yields

$$\sin\theta = \frac{\frac{\pi}{2}H_y \cos\eta}{\frac{1}{2}H_k \cos^2\eta - \frac{\pi}{2}H_D \cos\eta + H_\perp} \tag{12}$$

and

$$\sin\eta = \frac{\frac{\pi}{2}H_y \cos(\eta+\theta)}{H_k \cos\eta - \frac{\pi}{2}H_D}. \tag{13}$$

The solid lines in Figs. 3c,d show $m_x = \cos(\Phi)$ versus $H_x$ and $H_y$, which reproduces the gradual reversal (reduction) of $\chi$ under $H_x$ ($H_y$) observed experimentally. The parameters used here are $|H_D| = 2800$ Oe, with left-handed chirality, and $H_\perp = 6300$ Oe, close to the measured value. The tilt angles versus $H_x$, $H_y$ agree qualitatively with the full micromagnetic results, as seen in Figs. 4e,f. We note that the longitudinal field required to null $\chi$ significantly underestimates $H_D$ when DMI is strong, since DW tilting allows $\Phi$ to rotate more readily than if the DW normal remained fixed.

### III.e. Full Micromagnetic Treatment of Thiele Effective Forces

To treat domain canting and DW tilting simultaneously, we performed full 2D micromagnetic simulations (see Appendix B and Supplemental Material) of the equilibrium DW structure versus $H_x$ and $H_y$ (Fig. 4), and computed $\chi_{SHE}$ numerically from the Thiele expressions[45]

$$\vec{H}_{SHE}^{DW} = \frac{H_{SHE}}{2} \frac{1}{w} \int_{-\infty}^{\infty} \int_0^w \left[ (\hat{m} \times \hat{y}) \cdot \frac{\partial \hat{m}}{\partial x} \right] dy\, dx\, \hat{z} \equiv H_{SHE} I_{SHE} \hat{z}$$
$$\vec{H}_z^{DW} = \frac{H_z}{2} \frac{1}{w} \int_{-\infty}^{\infty} \int_0^w \left[ \hat{z} \cdot \frac{\partial \hat{m}}{\partial x} \right] dy\, dx\, \hat{z} \equiv H_z I_z \hat{z}$$
(14)

for the effective fields from the SHE and $H_z$, respectively. Under the usual 1D Walker ansatz for the DW structure, Eqs. (14) reduce to $H_{SHE}^{DW} = \frac{\pi}{2} H_{SHE} \cos\Phi$ and $H_z^{DW} = H_z$ so that $\chi_{SHE} = \frac{\pi}{2} \chi_{SHE}^0 \cos(\Phi)$ as expected. In the case of a general DW profile, the SHE acts like an easy-axis applied field $(I_{SHE}/I_z)H_{SHE}$, so that $\chi_{SHE} = (I_{SHE}/I_z)\chi_{SHE}^0$.

We used Eqs. (14) to fit the in-plane field-dependence of $\chi$ in Figs. 3c,d, by micromagnetically computing the DW structure as a function of $H_x$ and $H_y$ and numerically computing $\chi_{SHE} = (I_{SHE}/I_z)\chi_{SHE}^0$ using Eqs. (14). This fit used only two free parameters: the effective spin Hall angle $\theta_{SH}$, which determines the vertical scale factor, and the DMI exchange constant $D$.

We first determined $\theta_{SH} \approx +0.07$ from the value of $\chi$ measured at $H_x=H_y=0$ in Figs. 3c,d, which agrees well with $\theta_{SH}$ for Pt in Ref. 7. We then varied the single parameter $D$ to best match the field dependence of the normalized quantity $\chi/\chi(H_x = H_y = 0)$, while holding all other micromagnetic parameters fixed at their measured values. This one-parameter fit reproduces the experimental data remarkably well with the best-fit value $D$=-1.2 mJ/m$^2$ (solid circles in Figs. 3c,d).

There remains some discrepancy between data and fit when $H_x$ opposes $H_D$, which we attribute to local dispersion of $D$ due to interface disorder. The micromagnetic simulations in Fig.

4e predict a sudden onset of DW tilting at a critical $H_x$, where $\chi$ begins to drop (Fig. 3c). Local dispersion in $D$ would tend to broaden this transition by allowing the DW moment to rotate at lower $H_x$ in some regions due to locally smaller $D$. Refining the model to include dispersion is however beyond the present scope.

## IV. Discussion

The results in Fig. 3 show that DW motion in Pt/CoFe/MgO and Ta/CoFe/MgO can be accounted for quantitatively by the SHE and DMI, and that the variation of the current-induced effective field with in-plane applied fields provides a means to conveniently extract these parameters. In the case of weak DMI, a simple 1D model suffices for analyzing the experimental data, but in the case of strong DMI, where large in-plane fields are required to probe the stiffness of the homochiral Néel DWs, a numerical approach is required. Nonetheless, as shown above, Eqs. (14) provide a general framework to numerically fit the data in terms of just two free parameters, $\theta_{SH}$ and $D$, that are essentially uncorrelated.

We note that the calculations of $\chi$ used to fit the data in the analyses above are based on the equilibrium DW structure, whereas experiments are performed under conditions of thermally-activated DW propagation near the depinning threshold. Therefore, it could be expected that pinning could distort the DW and lead to deviations from the models used for fitting the data. However, the experimental analysis is based on measuring the average propagation field measured over many repeated propagation cycles across a relatively long propagation distance, thus probing the full ensemble of disorder and thermally-activated fluctuations. Deviations from the nominal DW line profile due to random distortions during propagation should hence average towards zero in the experimental determination of $\chi$. Indeed, we verified through finite-temperature micromagnetic simulations of DW propagation under realistic conditions of disorder[46] that the tilting predicted in Fig. 4 under static equilibrium is preserved on average during thermally-activated propagation in the presence of edge roughness, whereas in the absence of strong DMI no net tilting is observed.

The presence of a single strong pinning site at an edge could lead to preferential tilting of the DW in one direction that would be repetitive from cycle to cycle, systematically influencing the current-induced effective field. However, if the measured propagation field were due to a single dominant defect, the current-induced depinning efficiency would necessarily vary randomly and significantly from device to device, and from position to position along a given device, depending on the location and strength of this dominant, random pinning site. In the Supplemental Material, we show measurements of $H_{\text{prop}}$ versus $j_e$, for several nominally identical structures, for a range of structures with different widths, and measurements using different field-sweep frequencies (and hence propagation timescales). In all of these measurements the extracted $\chi$ is identical within experimental uncertainty, indicating that this parameter is a robust measure of the current-induced effective field.

Finally, we determined $H_{prop}$ and $j_{prop}$ through dynamical micromagnetic simulations of DW propagation with edge roughness, shown in detail in Supplemental Material, to verify that our quasistatic analysis of $\chi$ reproduces the full micromagnetics treatment. In these simulations, we used micromagnetic parameters for Pt/CoFe/MgO extracted from the analysis presented above, and included a random edge roughness with a characteristic grain size of $D_g = 4$nm. The propagation thresholds for field-driven and current-driven motion were determined separately, and their ratio used to determine $\chi$. With an effective spin Hall angle $\theta_{SH} = +0.07$ used in the simulations, the ratio $H_{prop}/j_{prop}$ (open diamonds in Fig. 3c) matches well with the experimentally measured $\chi$ and with $\chi$ calculated from the static micromagnetic DW structure via Eqs. (14) (solid points in Figs. 3c,d). These results further validate our approach to fitting the data numerically using the micromagnetically-computed equilibrium DW structure via Eqs. (14).

## V. Conclusions

The DMI constant $D$ takes the same sign for Pt/CoFe/MgO and Ta/CoFe/MgO but differs in magnitude by a factor of 20, while the spin Hall angle $\theta_{SH}$ alternates in sign from Pt to Ta but the magnitudes are within a factor of two. This suggests that the DMI and Slonczewski-like SOT, though related through SOC, derive from different microscopic mechanisms in these materials and can hence be independently tuned. In the case of strong DMI, the frequently used 1D model fails qualitatively to describe DW motion in the presence of large in-plane fields or strong torques on the DW moment. Both domain canting and DW tilting must be treated in full in order to quantitatively extract the DMI strength from experiments.

The DMI in Pt/CoFe/MgO is remarkably strong, comparable to that in ultrathin epitaxial layers grown on single crystal substrates.[26–28] This suggests the feasibility of realizing more complex spin textures[26,27,30,32] such as spin spirals and skyrmions in robust thin-film heterostructures. These should emerge for $|H_D|/H_\perp > 2/\pi$,[31] not far from $|H_D|/H_\perp \approx 0.45$ measured here for Pt/CoFe/MgO. The possibility to engineer spin torques and spin textures, using materials amenable to practical device integration, presents new opportunities for high-performance spintronics applications.

**APPENDIX A: Domain Wall Profile Under Large In-Plane Fields**

The DW profile in the presence of the in-plane longitudinal field was derived in Refs. 40–43 as

$$\theta(x,t) = \sin^{-1}\left(h + \frac{1-h^2}{\cosh[(x-q(t))/\Delta] + h}\right), \quad \Phi(x,t) = \Phi(t) \quad for \ x > q(t)$$
$$\theta(x,t) = \pi - \sin^{-1}\left(h + \frac{1-h^2}{\cosh[(x-q(t))/\Delta] + h}\right), \quad \Phi(x,t) = \Phi(t) \quad for \ x \leq q(t)$$
(A1)

where $\theta$ is the polar angle, $\Phi$ is the azimuthal angle, $q$ is the DW center position, $h \equiv H_x / H_\perp$ is the normalized external longitudinal field, and $H_\perp$ is the effective perpendicular anisotropy field. Here $\Delta$ is the DW width, defined as

$$\Delta = \frac{\sqrt{A/K_{u,eff}}}{\sqrt{(1+(K_s/K_{u,eff})\cos^2\Phi)(1-h^2)}}$$
(A2)

where $A$ is the exchange stiffness constant, $K_{u,eff} = \frac{1}{2}\mu_0 M_s H_\perp$ is the effective perpendicular anisotropy energy density, and $K_s = \frac{1}{2}\mu_0 M_s H_k$ is the DW (magnetostatic) shape anisotropy energy density. Eq. (A2) shows that the DW width depends on the in-plane field as well as the DW angle $\Phi$. Here we neglect its dependence on $\Phi$ for simplicity (by assuming that $K_s/K_{u,eff}$ is small), and apply the rigid DW approximation for a given $h$ in order to derive Thiele equations below. Within the rigid DW limit, we note that Eqs. (A1) and (A2) describe the DW profile regardless of the direction of in-plane field, so that these expressions are likewise applicable for transverse applied field by redefining $h \equiv H_y / H_\perp$. We obtain Thiele's equations from Eq. (A1) and (A2), as given by main text Eqs. (4)-(6).

Since the experiments are performed under quasistatic conditions of DW depinning and creep, we obtain $\Phi$ from the steady state solution ($\dot{q} = \dot{\Phi} = 0$) for the torque equation (Eq. (5) or Eq. (6)). In the case of $H_x$, the torque equation, Eq. (5), may be solved analytically, yielding

$$\cos\Phi = -\xi \frac{\pm D - \sqrt{1-h^2}\mu_0 H_x M_S \Delta}{K_s \Delta \sqrt{1-h^2}\left(\sqrt{1-h^2} + 2h\xi\right)}. \tag{A3}$$

Here, + (-) corresponds to up-down (down-up) DWs. Eq. (A3) can be re-written in terms of effective fields as

$$\cos\Phi = -\xi \frac{\pm H_D - \sqrt{1-h^2} H_x}{\frac{1}{2} H_k \sqrt{1-h^2}\left(\sqrt{1-h^2} + 2h\xi\right)}. \tag{A4}$$

Where the sign of $H_D$ alternates between up-down and down-up DWs. These expressions, restricted to the range $-1 \leq \cos\Phi \leq 1$, were used to generate the dotted curve in Fig. 3c in the main text.

We note that strictly speaking Eq. (A1) is analytically integrable to obtain the Thiele equations only in the case $|\cos\Phi| = 1$ corresponding to Néel DWs. Therefore, Eqs. (A3) and (A4) are not analytically exact solutions to the model. Nonetheless, these equations provide the range of $h$ over which the ansatz of a Néel DW holds, such that whenever $|\cos\Phi| > 1$ in these expressions, one fixes $\cos\Phi = \pm 1$ as appropriate and the derived Thiele equations are self-consistent with the assumed DW profile. The width of the transition regions depicted by the dotted curves in Fig. 3c is therefore accurate within this model, but the exact form of the transition will deviate from that predicted by Eqs. (A3) and (A4) and plotted in Fig. 3c. Importantly, the predicted variation of $\chi$ with $h$ in the field ranges where the DW is fully Néel is analytically well-motivated.

**APPENDIX B: Details of Micromagnetics Implementation**

*Energy and effective field* – The equilibrium states were computed by integrating the total energy density $\varepsilon$ over the sample $E = \int_V \varepsilon \, dV$. Apart from the standard exchange $\varepsilon_{exch}$, perpendicular magnetocrystalline anisotropy $\varepsilon_{ani,u}$ (uniaxial with easy-axis along the $z$ direction), magnetostatic $\varepsilon_{dmg}$ and external field $\varepsilon_{ext}$ contributions, it also accounts for the DMI $\varepsilon_{DMI}$. In the continuous approach for thin films (with dimensions $L_x$, $w$, $t$ along the Cartesian axes, and with $t \ll w, L_x$), the variations of the magnetization along the $z$-axis can be neglected, and the total energy density can be expressed as[31,47]

$$\begin{aligned}\varepsilon &= \varepsilon_{exch} + \varepsilon_{ani,u} + \varepsilon_{dmg} + \varepsilon_{ext} + \varepsilon_{DMI} = \\ &= A\left[\left(\frac{\partial \vec{m}}{\partial x}\right)^2 + \left(\frac{\partial \vec{m}}{\partial y}\right)^2\right] + K_u\left(1 - (\vec{m}\cdot u_k)^2\right) - \\ &\quad - \frac{1}{2}\mu_0 M_s \vec{H}_{dmg}\cdot\vec{m} - \mu_0 M_s \vec{H}_{ext}\cdot\vec{m} + \\ &\quad + D\left[\left(m_z\frac{\partial m_x}{\partial x} - m_x\frac{\partial m_z}{\partial x}\right) + \left(m_z\frac{\partial m_y}{\partial y} - m_y\frac{\partial m_z}{\partial y}\right)\right]\end{aligned} \quad (B1)$$

where $(m_x, m_y, m_z)$ are the local Cartesian components of the magnetization vector $\vec{m}(\vec{r}) = \vec{M}(\vec{r})/M_s$ normalized to the saturation magnetization $M_s$, $A$ is the exchange constant, $K_u$ is the perpendicular magnetocrystalline anisotropy constant and $D$ is the DMI parameter. $\vec{H}_{dmg}(\vec{r})$ is the magnetostatic field computed from the magnetization distribution through the magnetostatics equations, and $\vec{H}_{ext} = (H_x, H_y, 0)$ is the externally applied in-plane magnetic field.

Similar to conventional micromagnetic formalism,[48] the static equilibrium state can be obtained from the calculus of variations ($\delta E = 0$) and expressed as a zero-torque condition in terms of a local effective field $\vec{H}_{eff}(\vec{r})$, as

$$\vec{m}(\vec{r}) \times \vec{H}_{eff}(\vec{r}) = 0 \qquad (B2)$$

for each point $\vec{r}$ of the sample, where the local effective field $\vec{H}_{eff}(\vec{r})$ is

$$\vec{H}_{eff}(\vec{r}) = -\frac{1}{\mu_0 M_s}\frac{\delta \varepsilon}{\delta \vec{m}} = \frac{2A}{\mu_0 M_s}\nabla^2 \vec{m} + \frac{2K_u}{\mu_0 M_s} m_z \vec{u}_z + \vec{H}_{dmg} + \vec{H}_{ext}$$
$$+ \frac{2D}{\mu_0 M_s}\left[\frac{\partial m_z}{\partial x}\vec{u}_x + \frac{\partial m_z}{\partial y}\vec{u}_y - \left(\frac{\partial m_x}{\partial x} + \frac{\partial m_y}{\partial y}\right)\vec{u}_z\right] \qquad (B3)$$

The magnetostatic field $\vec{H}_{dmg}(\vec{r})$ is evaluated by means of the Fast Fourier Transform (FFT) and the zero padding technique using the Newell's expressions for the magnetostatics.[49] See Ref. 44 for further details.

*Boundary conditions* – In the absence of DMI, the exchange interaction imposes boundary conditions at the surfaces of the sample[50] such that the magnetization vector does not change along the surface normal $\vec{n}$, that is

$$\frac{\partial \vec{m}}{\partial n} = 0 \qquad (B4)$$

where $\partial/\partial n$ indicates the derivative in the outside direction normal to the surface of the sample. However, in the presence of the DMI, this boundary condition has to be replaced by

$$\frac{d\vec{m}}{dn} = \frac{D}{2A}\vec{m} \times (\vec{n} \times \vec{u}_z) \qquad (B5)$$

*Solver* – The sample was discretized using a 2D mesh with a lateral cell size of 4 nm. For each applied in-plane field, Eq. (B2) together with Eqs. (B3) and (B5) were iteratively solved by means of a Conjugate Gradient solver.[51] The equilibrium state is assumed to be achieved when this condition is reached with a maximum error of $|\vec{m}(\vec{r}) \times \vec{H}_{eff}(\vec{r})| < 10^{-5}$ for all computational cells.

*Simulation of domain wall displacement* – Domain wall motion assisted by a spatially uniform current density along the *x*-axis $\vec{j}_a = j_a \vec{u}_x$, is studied by solving the augmented Landau-Lifshitz-Gilbert equation[23,31]

$$\frac{d\vec{m}}{dt} = -\gamma_0 \vec{m} \times \vec{H}_{eff} + \alpha \vec{m} \times \frac{d\vec{m}}{dt} + \gamma_0 \frac{\hbar \theta_{SH} j_a}{2e\mu_0 M_s L_z} \vec{m} \times (\vec{m} \times \vec{u}_y) \quad , \tag{B6}$$

where $\gamma_0$ is the gyromagnetic ratio, $\hbar$ is the reduced Planck constant, $e$ is the electric charge and $\mu_0$ is the permeability of free space. The first term on the right hand side of Eq. (B6) describes the local magnetization precession around the local effective field $\vec{H}_{eff}$, which includes exchange, magnetostatic, uniaxial perpendicular magnetocrystalline anisotropy, external field $\vec{H}_{ext} = (H_x, H_y, H_z)$ and DMI contributions as described above. The second term accounts for the dissipation with the dimensionless Gilbert damping parameter set to $\alpha = 0.3$. The last term on the right hand side of Eq. (B6) is the Slonczewski-like torque due to the SHE with $\theta_{SH} = +0.07$. The sample was discretized using a 2D mesh with a lateral cell size of 4 nm, and Eq. (B6) was numerically solved by means of a 4$^{th}$ Runge-Kutta algorithm with a time step of 65 fs.

**Acknowledgements**

This work was supported in part by the National Science Foundation under NSF-ECCS-1128439. Devices were fabricated using instruments in the MIT Nanostructures Laboratory, the Scanning Electron-Beam Lithography facility at the Research Laboratory of Electronics, and the Center for Materials Science and Engineering at MIT. S.E. acknowledges financial support by the NSF Graduate Research Fellowship Program. The work by E.M. was supported by projects MAT2011-28532-C03-01 from the Spanish government and SA163A12 from Junta de Castilla y Leon. The work by K.J.L. was supported by the NRF (NRF-2013R1A2A2A01013188).

**References**

[1] T.A. Moore, I.M. Miron, G. Gaudin, G. Serret, S. Auffret, B. Rodmacq, A. Schuhl, S. Pizzini, J. Vogel, and M. Bonfim, Appl. Phys. Lett. **93**, 262504 (2008).
[2] I.M. Miron, P.-J. Zermatten, G. Gaudin, S. Auffret, B. Rodmacq, and A. Schuhl, Phys. Rev. Lett. **102**, 137202 (2009).
[3] I.M. Miron, G. Gaudin, S. Auffret, B. Rodmacq, A. Schuhl, S. Pizzini, J. Vogel, and P. Gambardella, Nat. Mater. **9**, 230 (2010).
[4] U.H. Pi, K. Won Kim, J.Y. Bae, S.C. Lee, Y.J. Cho, K.S. Kim, and S. Seo, Appl. Phys. Lett. **97**, 162507 (2010).
[5] I.M. Miron, T. Moore, H. Szambolics, L.D. Buda-Prejbeanu, S. Auffret, B. Rodmacq, S. Pizzini, J. Vogel, M. Bonfim, A. Schuhl, and G. Gaudin, Nat. Mater. **10**, 419 (2011).
[6] I.M. Miron, K. Garello, G. Gaudin, P.-J. Zermatten, M.V. Costache, S. Auffret, S. Bandiera, B. Rodmacq, A. Schuhl, and P. Gambardella, Nature **476**, 189 (2011).
[7] L. Liu, O.J. Lee, T.J. Gudmundsen, D.C. Ralph, and R.A. Buhrman, Phys. Rev. Lett. **109**, 096602 (2012).
[8] L. Liu, C.-F. Pai, Y. Li, H.W. Tseng, D.C. Ralph, and R.A. Buhrman, Science **336**, 555 (2012).
[9] C.-F. Pai, L. Liu, Y. Li, H.W. Tseng, D.C. Ralph, and R.A. Buhrman, Appl. Phys. Lett. **101**, 122404 (2012).
[10] S. Emori, D.C. Bono, and G.S.D. Beach, Appl. Phys. Lett. **101**, 042405 (2012).
[11] K.-S. Ryu, L. Thomas, S.-H. Yang, and S.S.P. Parkin, Appl. Phys. Express **5**, 093006 (2012).
[12] J. Kim, J. Sinha, M. Hayashi, M. Yamanouchi, S. Fukami, T. Suzuki, S. Mitani, and H. Ohno, Nat. Mater. (2012).
[13] X. Fan, J. Wu, Y. Chen, M.J. Jerry, H. Zhang, and J.Q. Xiao, Nat. Commun. **4**, 1799 (2013).
[14] P.P.J. Haazen, E. Murè, J.H. Franken, R. Lavrijsen, H.J.M. Swagten, and B. Koopmans, Nat. Mater. **12**, 299 (2013).
[15] S. Emori, U. Bauer, S.-M. Ahn, E. Martinez, and G.S.D. Beach, Nat. Mater. **12**, 611 (2013).
[16] K.-S. Ryu, L. Thomas, S.-H. Yang, and S. Parkin, Nat. Nanotechnol. **8**, 527 (2013).
[17] T. Koyama, H. Hata, K.-J. Kim, T. Moriyama, H. Tanigawa, T. Suzuki, Y. Nakatani, D. Chiba, and T. Ono, Appl. Phys. Express **6**, 033001 (2013).
[18] K. Garello, I.M. Miron, C.O. Avci, F. Freimuth, Y. Mokrousov, S. Blügel, S. Auffret, O. Boulle, G. Gaudin, and P. Gambardella, Nat. Nanotechnol. **advance online publication**, (2013).
[19] X. Wang and A. Manchon, Phys. Rev. Lett. **108**, 117201 (2012).
[20] K.-W. Kim, S.-M. Seo, J. Ryu, K.-J. Lee, and H.-W. Lee, Phys. Rev. B **85**, 180404 (2012).
[21] P.M. Haney, H.-W. Lee, K.-J. Lee, A. Manchon, and M.D. Stiles, Phys. Rev. B **87**, 174411 (2013).
[22] E. Martinez, J. Appl. Phys. **111**, 033901 (2012).
[23] E. Martinez and G. Finocchio, IEEE Trans. Magn. **49**, 3105 (2013).
[24] A. Brataas, A.D. Kent, and H. Ohno, Nat. Mater. **11**, 372 (2012).
[25] A. Thiaville, Y. Nakatani, J. Miltat, and Y. Suzuki, Europhys. Lett. EPL **69**, 990 (2005).
[26] M. Bode, M. Heide, K. von Bergmann, P. Ferriani, S. Heinze, G. Bihlmayer, A. Kubetzka, O. Pietzsch, S. Blügel, and R. Wiesendanger, Nature **447**, 190 (2007).
[27] S. Heinze, K. von Bergmann, M. Menzel, J. Brede, A. Kubetzka, R. Wiesendanger, G. Bihlmayer, and S. Blügel, Nat. Phys. **7**, 713 (2011).

[28] M. Heide, G. Bihlmayer, and S. Blügel, Phys. Rev. B **78**, 140403 (2008).

[29] G. Chen, J. Zhu, A. Quesada, J. Li, A.T. N'Diaye, Y. Huo, T.P. Ma, Y. Chen, H.Y. Kwon, C. Won, Z.Q. Qiu, A.K. Schmid, and Y.Z. Wu, Phys. Rev. Lett. **110**, 177204 (2013).

[30] A. Fert, V. Cros, and J. Sampaio, Nat. Nanotechnol. **8**, 152 (2013).

[31] A. Thiaville, S. Rohart, É. Jué, V. Cros, and A. Fert, EPL Europhys. Lett. **100**, 57002 (2012).

[32] O.A. Tretiakov and A. Abanov, Phys. Rev. Lett. **105**, 157201 (2010).

[33] Current densities were estimated by assuming current flow only through the CoFe layer and the adjacent heavy metal layer, so that the effective conductive thickness was 5.6 nm for Ta/CoFe/MgO and 3.6 nm for Pt/CoFe/MgO. We neglected current shunting in the bottom Ta seed layer in the Pt/CoFe/MgO, as sputtered Ta is at least 5 times more resistive than Pt.

[34] We estimate the misalignment of the in-plane field to be less than ~2°. The out-of-plane component of the field misalignment was calibrated through the offset in the DW propagation field in the nanostrip or the nucleation field of a nearby 20 μm square film.

[35] L. San Emeterio Alvarez, K.-Y. Wang, S. Lepadatu, S. Landi, S.J. Bending, and C.H. Marrows, Phys. Rev. Lett. **104**, 137205 (2010).

[36] J.-C. Lee, K.-J. Kim, J. Ryu, K.-W. Moon, S.-J. Yun, G.-H. Gim, K.-S. Lee, K.-H. Shin, H.-W. Lee, and S.-B. Choe, Phys. Rev. Lett. **107**, 067201 (2011).

[37] A.V. Khvalkovskiy, V. Cros, D. Apalkov, V. Nikitin, M. Krounbi, K.A. Zvezdin, A. Anane, J. Grollier, and A. Fert, Phys. Rev. B **87**, 020402 (2013).

[38] T. Koyama, D. Chiba, K. Ueda, K. Kondou, H. Tanigawa, S. Fukami, T. Suzuki, N. Ohshima, N. Ishiwata, Y. Nakatani, K. Kobayashi, and T. Ono, Nat. Mater. **10**, 194 (2011).

[39] S.V. Tarasenko, A. Stankiewicz, V.V. Tarasenko, and J. Ferré, J. Magn. Magn. Mater. **189**, 19 (1998).

[40] J. Kaczér and R. Gemperle, Czech J Phys B **11**, 510 (1961).

[41] V.L. Sobolev, H.L. Huang, and S.C. Chen, J. Magn. Magn. Mater. **147**, 284 (1995).

[42] O. Boulle, L.D. Buda-Prejbeanu, M. Miron, and G. Gaudin, J. Appl. Phys. **112**, 053901 (2012).

[43] O. Boulle, L.D. Buda-Prejbeanu, E. Jué, I.M. Miron, and G. Gaudin, J. Appl. Phys. **115**, 17D502 (2014).

[44] L. Lopez-Diaz, D. Aurelio, L. Torres, E. Martinez, M.A. Hernandez-Lopez, J. Gomez, O. Alejos, M. Carpentieri, G. Finocchio, and G. Consolo, J. Phys. Appl. Phys. **45**, 323001 (2012).

[45] A.A. Thiele, Phys. Rev. Lett. **30**, 230 (1973).

[46] E. Martinez, S. Emori, N. Perez, L. Torres, and G.S.D. Beach, J. Appl. Phys. **115**, 213909 (2014).

[47] A. Hubert and R. Schafer, *Magnetic Domains: The Analysis of Magnetic Microstructures* (Springer, Berlin; New York, 1998).

[48] W.F. Brown, *Micromagnetics* (Interscience Publishers, 1963).

[49] A.J. Newell, W. Williams, and D.J. Dunlop, J. Geophys. Res. Solid Earth **98**, 9551 (1993).

[50] G. Bertotti, *Hysteresis in Magnetism: For Physicists, Materials Scientists and Engineers* (Acad. Press, 1998).

[51] W.H. Press, *Numerical Recipes in Fortran 90: The Art of Parallel Scientific Computing* (Cambridge University Press, New York, 1996).

[52] While this manuscript was under review, a theoretical treatment predicting DW tilting was independently published in this journal as Phys. Rev. Lett. **111,** 217203 (2013).

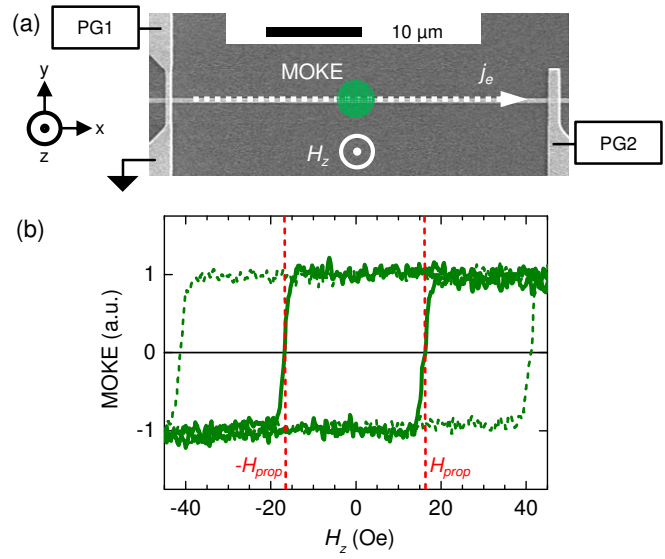

FIG. 1 (color online) (a) Illustration of experimental setup superposed on micrograph of a nanostrip device. Pulse Generator 1 (PG1) outputs the DW nucleation pulse and PG 2 outputs $j_e$ along nanostrip. (b) Polar MOKE hysteresis loops obtained with DWs initialized by the nucleation pulse (solid curve) and without nucleation pulses (dotted curve). $H_{prop}$ is indicated by red dotted lines.

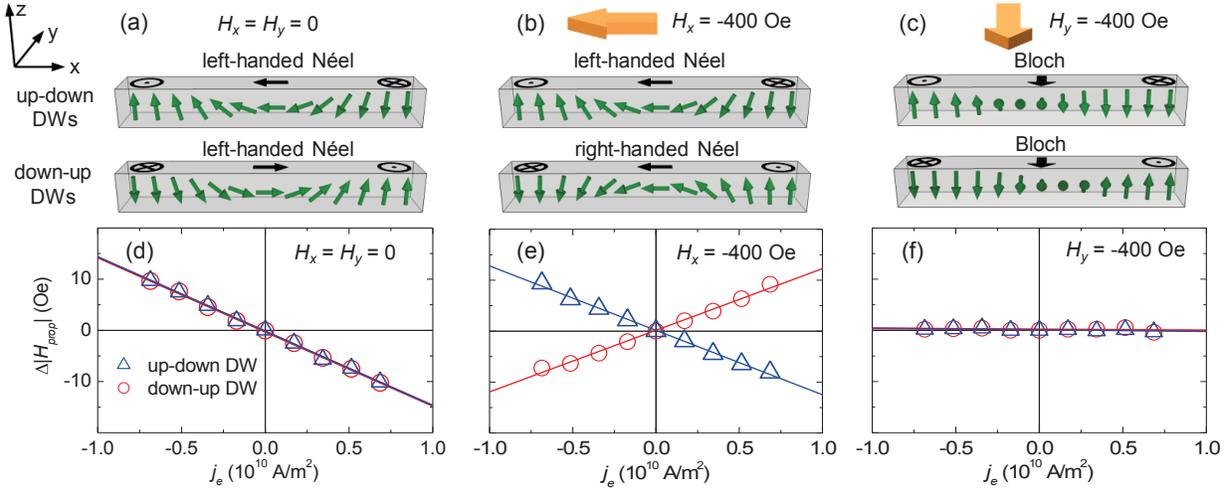

FIG. 2 (color online) (a-c) Illustrations of DWs under (a) zero in-plane field, (b) longitudinal field $H_x$, and (c) transverse field $H_y$. (d-f) Change in $H_{prop}$ versus $j_e$ in Ta/CoFe/MgO corresponding to (a-c).

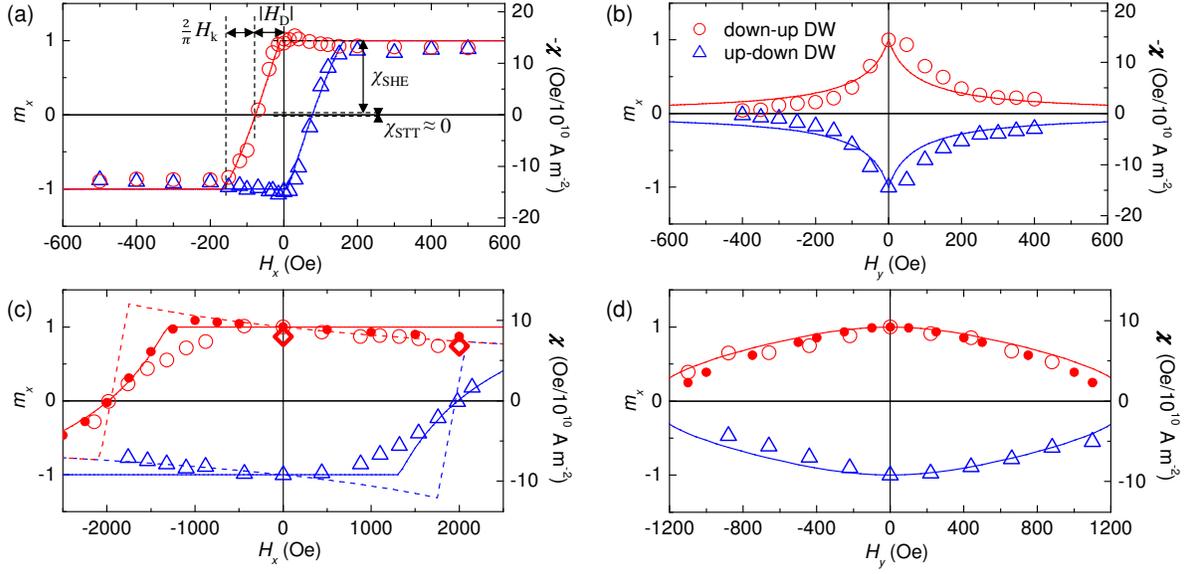

FIG. 3 (color online) Measured $\chi$ (open symbols), analytically computed longitudinal component $m_x$ of DW moment (solid curves) and micromagnetically computed $\chi$ (solid circles). (a,b) Ta/CoFe/MgO under (a) $H_x$ and (b) $H_y$; $-\chi$ plotted to account for negative spin Hall angle. (c,d) Pt/CoFe/MgO under (c) $H_x$ and (d) $H_y$. Dotted curves in (c) show analytical correction to $\chi$ due to domain canting for $H_x$ parallel to $H_D$. Open diamonds in (c) denote $\chi$ obtained from micromagnetic simulations of field and current-driven DW propagation with edge roughness; uncertainty is of order the symbol size.

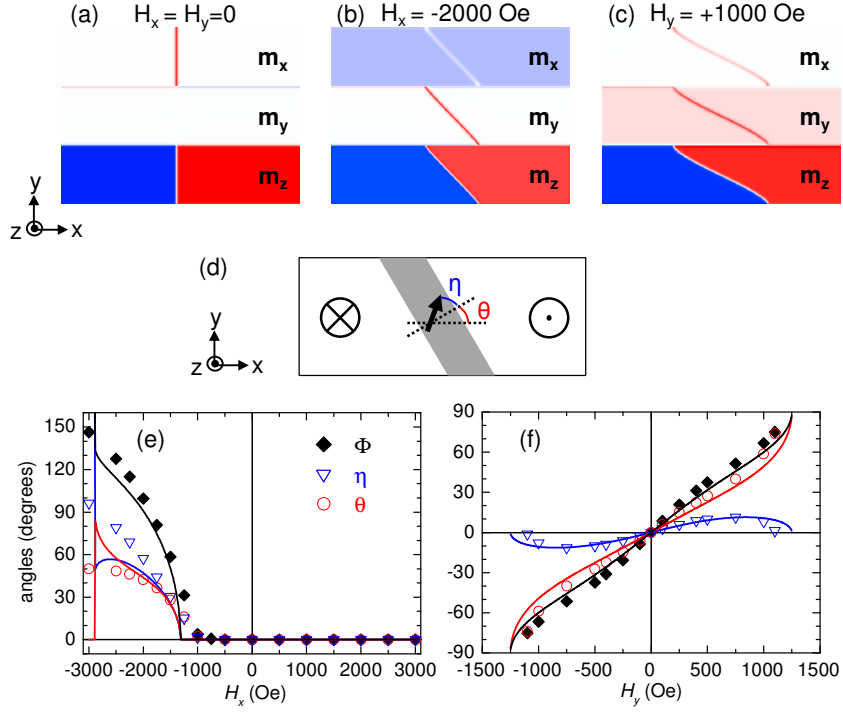

FIG. 4 (color online) (a-c) Micromagnetic snapshots of down-up DWs under (a) zero in-plane field, (b) $H_x$ against the DMI, and (c) $H_y$. The three Cartesian components of $m$ are shown, with red=+1, white=0, and blue =-1. (d) Definitions of angles $\eta$ and $\theta$. (e,f) DW rotation parameterized by $\Phi=\eta+\theta$ under (e) $H_x$ and (f) $H_y$ for micromagnetic (symbols) and analytical (lines) calculations.

Supplemental Material:

Spin Hall torque magnetometry of Dzyaloshinskii domain walls

Satoru Emori[1], Eduardo Martinez[2], Kyung-Jin Lee[3,4], Hyun-Woo Lee[5],

Uwe Bauer[1], Sung-Min Ahn[1], Parnika Agrawal[1], David C. Bono[1], and Geoffrey S. D. Beach[1*]

[1] Department of Materials Science and Engineering, Massachusetts Institute of Technology,

Cambridge, Massachusetts 02139, USA

[2] Dpto. Física Aplicada. Universidad de Salamanca,

Plaza de los Caidos s/n E-38008, Salamanca, Spain

[3] Department of Materials Science and Engineering, Korea University, Seoul 136-701, Korea

[4] KU-KIST Graduate School of Converging Science and Technology, Korea University, Seoul

136-713, Korea

[5] PCTP and Department of Physics, Pohang University of Science and Technology, Kyungbuk

790-784, Korea

## I. Domain wall propagation field governed by fine-scale defects

Time-of-flight domain wall (DW) velocity measurements (Fig. S1 and Refs. [1,2]) indicate a uniform average DW velocity along the strip under a constant driving current. For example, even at a low current density $j_e \sim 10^{10}$ A/m$^2$, DWs moved uniformly on average at ~0.01 m/s (Figs. S1(b),(c)). Plots of the logarithm of the DW velocity against $j_e^{-1/4}$ and $H_z^{-1/4}$ (Fig. S2) yield linear relationships, suggesting DW motion is well described by two-dimensional creep scaling [3,4]. Therefore, for the results shown in this study, $H_{prop}$ is governed by DW pinning from nanoscale inhomogeneity (e.g. film roughness, grain boundaries, etc.) distributed throughout each nanostrip, rather than from a single dominant defect in a nanostrip. This latter case would lead to discontinuities in the position versus time measurements, which is not observed. The typical $H_{prop}$ in the absence of $j_e$ and in-plane bias fields was ~20 Oe in Ta/CoFe/MgO and Pt/CoFe/MgO nanostrips.

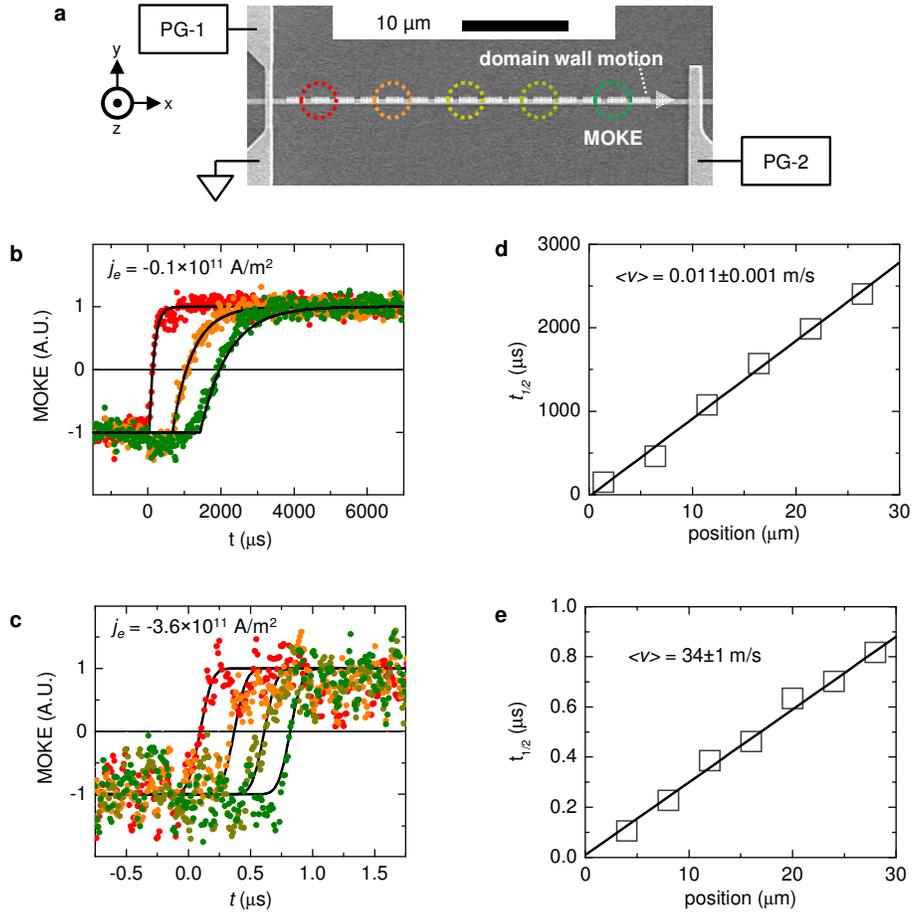

**Figure S1.** (a) Schematic of the time-of-flight DW motion measurement. The MOKE laser spot is placed at several positions along the nanostrip. At each position, an averaged MOKE transient (magnetization switching due to DW switching as a function of time) is measured. (b,c) Normalized MOKE transients at different positions in a Pt/CoFe/MgO strip at (b) a small driving current and (c) large driving current. (d,e) DW arrival time $t_{1/2}$ (time at which the zero-crossing of the normalized MOKE transient occurs) plotted against measured position. (d) corresponds to the small-current case (in (c)), and (e) to the large-current case (in (d)).

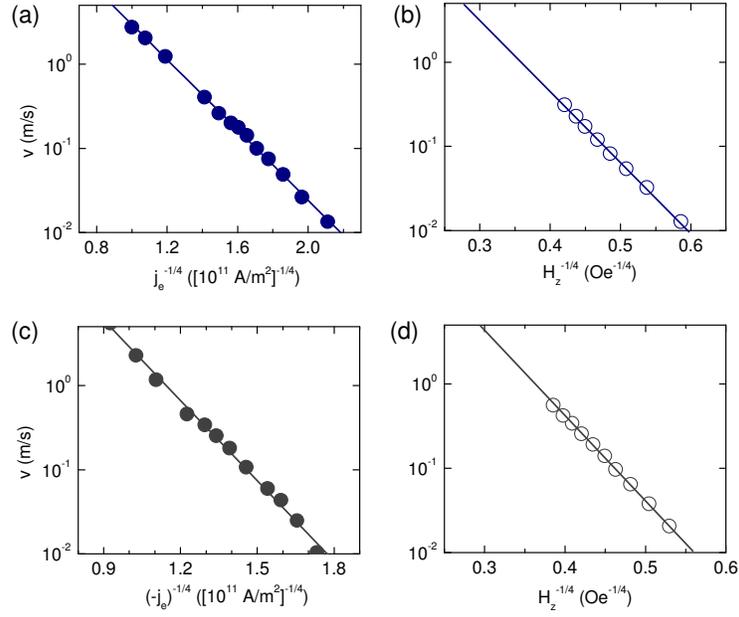

**Figure S2.** DW velocity plotted against $j_e^{-1/4}$ and $H_z^{-1/4}$ for (a,b) Ta/CoFe/MgO and (c,d) Pt/CoFe/MgO. The lower maximum measured velocity driven by $H_z$ (b,d) is due to random domain nucleation.

Because $H_{prop}$ is a measure of thermally activated DW motion, the baseline $H_{prop}$ at $j_e = 0$ may vary from sample to sample (Figs. S3(a),(b) and S4(a)) or with the sweep rate of the driving field $H_z$ (Fig. S4(b)). $H_{prop}$ is modified by a constant current $j_e \neq 0$ injected during the $H_z$ sweep. The change in $H_{prop}$ scales linearly with the value of $j_e$, as shown in Figs. S3 and S4, indicating that $j_e$ is equivalent to a DC offset in $H_z$ that drives DWs. The slope $\chi = \Delta H_{prop}/\Delta j_e$ is essentially the same (varying at most by ~10%) for different samples and sweep rates (Figs. S3 and S4). Thus, $\chi$ is a robust measure of the current-induced effective field, from which we can quantify the spin Hall effect and the x-component of the DW magnetic moment $m_x$.

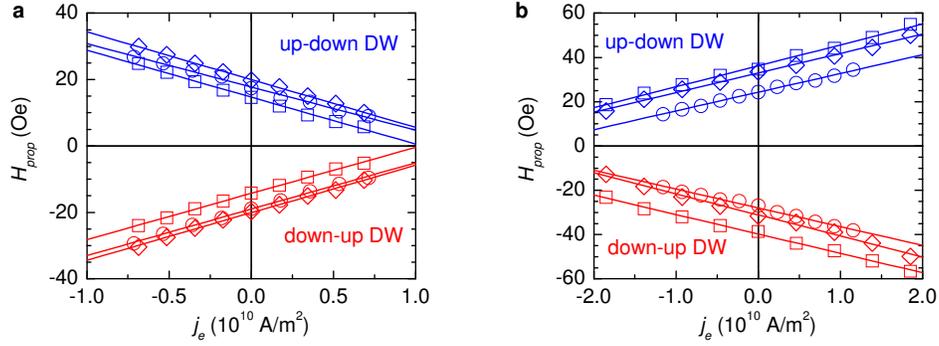

**Figure S3.** Change in the DW propagation field $H_{prop}$ with respect to electron current density $j_e$ in nominally identical samples of (a) Ta/CoFe/MgO and (b) Pt/CoFe/MgO.

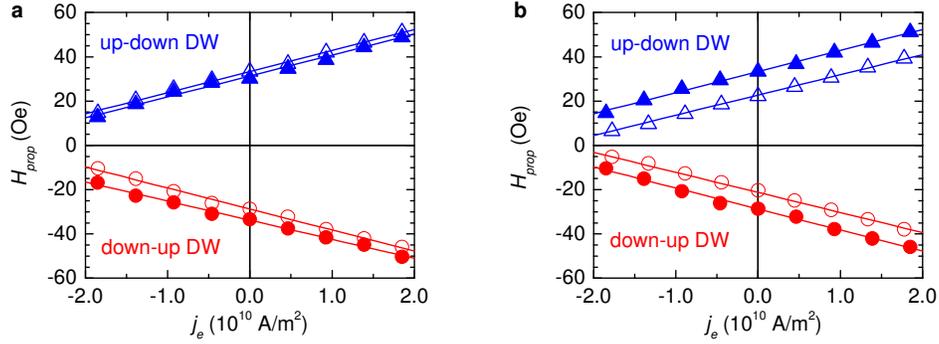

**Figure S4.** Change in the DW propagation field $H_{prop}$ with respect to electron current density $j_e$ in Pt/CoFe/MgO nanostrips (a) with different widths (empty symbol: 500 nm; filled symbol: 1200 nm), and (b) under different field sweep rates and MOKE laser positions (empty symbol: ≈10 Oe/ms, ≈20 µm away from the nucleation line; filled symbol: ≈500 Oe/ms, ≈5 µm away from the nucleation line).

In-plane fields ($H_x$ and $H_y$) cant the domain magnetization away from the out-of-plane easy axis. With increasing $H_x$ or $H_y$, the nucleation field $H_{nuc}$ decreases, thereby making it more difficult to isolate DW propagation from random nucleation of reverse domains (i.e., $H_{nuc}$ approaches $H_{prop}$). This was especially problematic in measurements of Pt/CoFe/MgO, for which large in-plane fields were required to produce a considerable change in $\chi$ (or $m_x$). The results shown in Figs. 3 (c),(d) of the main text were obtained from 1200-nm wide Pt/CoFe/MgO strips, which typically had larger $H_{nuc}$ than 500-nm wide strips. $H_{nuc}$ also remained greater than the $H_{prop}$ over a greater range of in-plane fields by conducting these measurements with a faster field sweep rate ≈500 Oe/ms (compared to ≈10 Oe/ms used for Ta/CoFe/MgO) and the MOKE laser placed ≈5 µm away from the nucleation line (compared to ≈20 µm for Ta/CoFe/MgO). The difference in the sample width or the measurement parameters did not affect the slope $\chi$, as shown in Fig. S4.

## II. Micromagnetic simulation of DW propagation field and propagation current

Micromagnetic simulations were performed to evaluate the propagation field $H_{prop}$ and the propagation current $j_{prop}$, and from them the field-to-current correspondence (efficiency) $\chi$ in the presence of in-plane longitudinal field parallel to the equilibrium DMI-stabilized internal DW moment. The disorder to impede DW motion was incorporated with an edge roughness with a typical grain size of $D_g$ = 4 nm on both sides of the strip (see [5] for further details). Such random disorder is qualitatively consistent with nanoscale defects distributed throughout experimentally measured strips (as described in Sec. I). The dimensions of the computed sample are length $L_x$ = 2800 nm, width $w$ = 160 nm and thickness $t$ = 0.6nm, with material parameters $M_s$ = 7×10$^5$ A/m; $K_u$ = 4.8×10$^5$ J/m$^3$; $A$ = 10$^{-11}$ J/m; $\alpha$ = 0.3; and $\theta_{SH}$ ≈ +0.07. The smaller strip width was chosen to save computational time

The "propagation values" $H_{prop}$ and $j_{prop}$ are defined as the minimum field (along $z$-axis) and the minimum current density (along the $x$-axis) required to promote sustained DW motion along a distance of 1.2µm. Below these threshold values, the DW displaces some distance from its initial position until reaching a final position where it remains pinned. The present study was performed for an up-down wall with left handed chirality, so that the internal DW moment points along the negative x-axis at rest. The aims of this study are to

1) Verify that the efficiency $\chi$ determined from Thiele effective forces computed for the equilibrium micromagnetic DW configuration agrees with the value determined from micromagnetically simulated DW propagation (depinning) in the presence of disorder (which might distort the DWs and change the efficiency)

2) Verify micromagnetically the decrease of the DW efficiency observed in the experimental measurements and predicted from Thiele force analysis of equilibrium DW structures, when an in-plane longitudinal field $H_x$ is applied in the same direction as the equilibrium DW moment (preferred by DMI).

We used micromagnetic parameters corresponding to the Pt/CoFe/MgO sample, and simulated a strip with random edge roughness with a characteristic grain size of $D_g = 4$nm. We considered purely field-driven motion, and purely current-driven motion separately, and varied field (current) in steps of 0.5 Oe ($0.005 \times 10^{12}$ A/m$^2$) near the depinning threshold in order to determine $H_{prop}$ and $j_{prop}$ respectively. Figure S5 shows representative DW position versus time curves that illustrate the behaviour just below and just above the (zero-temperature) depinning thresholds.

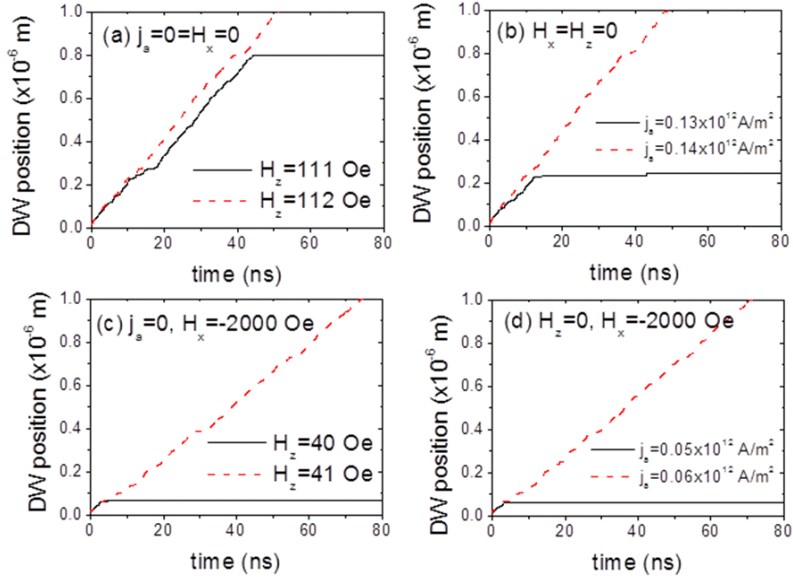

**Figure S5.** (a-b) Micromagnetically computed DW position versus time under $H_z$ or $j_a$, both at $H_x=0$ and $H_x=-2000$ Oe oriented parallel to the DMI-stabilized DW moment.

The efficiency $\chi$ is defined in the same manner as in the experiments, that is the ratio of $H_{prop}$ to $j_{prop}$ which yields

$$\chi_{\mu M}(H_x = 0; D_g = 4nm; T = 0) = \frac{H_{prop}(j_a = H_z = 0)}{j_{prop}(H_z = H_x = 0)} \approx \frac{112 \text{Oe}}{0.14 \times 10^{12} \text{A/m}^2} = \frac{8.0 \text{ Oe}}{10^{10} \text{A/m}^2}$$

for $H_x=0$, and

$$\chi_{\mu M}(H_x = -2000\text{Oe}; D_g = 4nm; T = 0) = \frac{H_{prop}(j_a = 0; H_x = -2000\text{Oe})}{j_{prop}(H_z = 0; H_x = -2000\text{Oe})}$$

$$\approx \frac{41 \text{Oe}}{0.06 \times 10^{12} \text{A/m}^2} = \frac{6.8 \text{ Oe}}{10^{10} \text{A/m}^2}$$

for $H_x=-2000$ Oe.

This value is around a 15% smaller than in the absence of in-plane field, and it is also in good agreement with experimental observations and with the micromagnetically-computed efficiency based on numerical analysis of the equilibrium DW configurations presented in the main text. The absolute values of $\chi$ agree with the effective field expected from a spin Hall angle of +0.07 used in the simulations. These micromagnetic results are depicted by open diamonds in Fig. 3(c) of the main text.

We note that in addition to the slight reduction of $\chi$ under $H_x$, explained above analytically due to the domain canting effect, there is a substantial reduction in the absolute pinning strength (i.e., both $H_{prop}$ and $j_{prop}$ are significantly reduced under large $H_x$, compared to the $H_x=0$ case). This effect arises from the variation in the DW energy density under in-plane fields, which was computed analytically and applied to the case of DW creep in Ref. [6].

### III. Asymmetry in $\chi$ under $H_y$ in Ta/CoFe/MgO

In Fig. 3 of the main text, the current-induced effective z-axis field $H_z^{eff} = \chi j_e$, was extracted from the slope of the propagation field versus current (its sign determined by considering the direction that up-down or down-up DWs are driven by current). The data show $\chi$ is asymmetric with respect to $H_y$ in Ta/CoFe/MgO (Fig. 3(b) in the main text). For both up-down and down-up DWs, the decrease in $\chi$ is larger for $H_y < 0$.

We verified that this asymmetry does not arise from misalignment of $H_y$, by measuring $\chi$ (defined here as the slope of $H_{prop}$ versus $j_e$) under various nominal field misalignments $\delta$. As shown in Fig. S6(c), the intentional misalignment does not eliminate the asymmetry. However, $\chi$ changes differently for up-down and down-up DWs under field misalignment, e.g. for $H_y > 0$ and $\delta = 6°$, it is clear that $\chi$ increases for up-down DWs whereas it decreases for down-up DWs. For up-down DWs (Fig. S6(a)), the longitudinal (-x) component of misaligned $H_y > 0$ aligns the internal moment closer to the DMI-stabilized -x orientation, so that the efficiency to the spin Hall effect does not decrease as much. By contrast, for the down-up DW (Fig. S6(b)), the same misaligned $H_y > 0$ rotates the moment farther away from the DMI-stabilized $+x$ orientation, thereby reducing the spin Hall torque efficiency more than for the up-down DW.

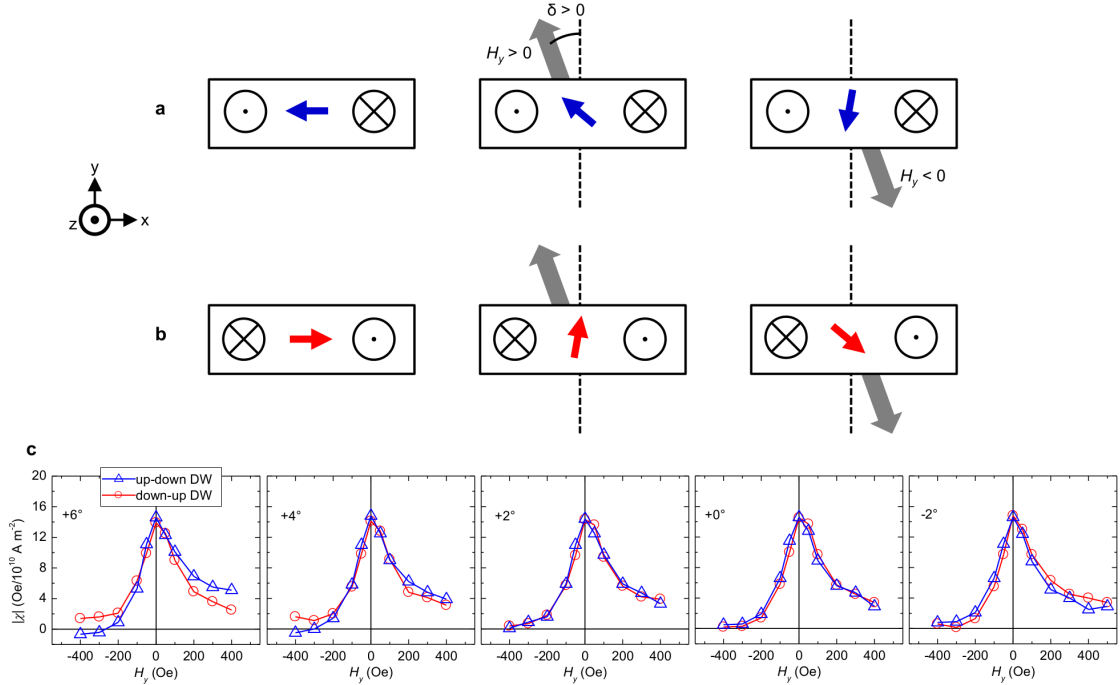

**Figure S6.** (a-b) Illustrations of the internal moment orientation for the (a) up-down DW and (b) down-up DW under transverse field $H_y$ with a misalignment $\delta$. (c) Efficiency $\chi$ versus $H_y$ at different misalignments $\delta$.

The current-induced Oersted field or the transverse field from the Rashba effect may be expected to play a role in the asymmetry under $H_y$. In particular, previous studies have reported large apparent transverse-field-like torques in Ta/CoFe(B)/MgO (Refs. [1,7,8]), which could arise from the transverse Rashba field $H_R$. (By contrast, in Pt/CoFe/MgO, the Rashba-like field is negligibly small [1].) $H_R$ scales linearly with electron current density $j_e$, and the direction of $H_R$ reverses if $j_e$ is reversed. Therefore, under a fixed $H_y$, the transverse $H_R$ should enhance or hinder the rotation of the DW moment with $H_y$, depending on the direction of $j_e$ (Figs. S7(a),(b)).

This would lead to a nonlinear relation between $\Delta H_{prop}$ and $j_e$ (Figs. S7(c),(d)), in which the slope (efficiency $\chi$) increases with larger $|j_e|$ when $H_R$ and $H_y$ are antiparallel, and decreases when $H_R$ and $H_y$ are parallel. Because $|j_e|$ was small at ~$10^{10}$ A m$^{-2}$ in our measurements, this nonlinearity was negligible, and $\Delta H_{prop}$ versus $j_e$ could be fit linearly. The linear slopes, and hence $\chi$, would be the same for $H_y > 0$ and $H_y < 0$, as illustrated in a schematic representation of the effect $H_R$ would have on $\Delta H_{prop}$ versus $j_e$ shown in Fig. S7(c),(d).

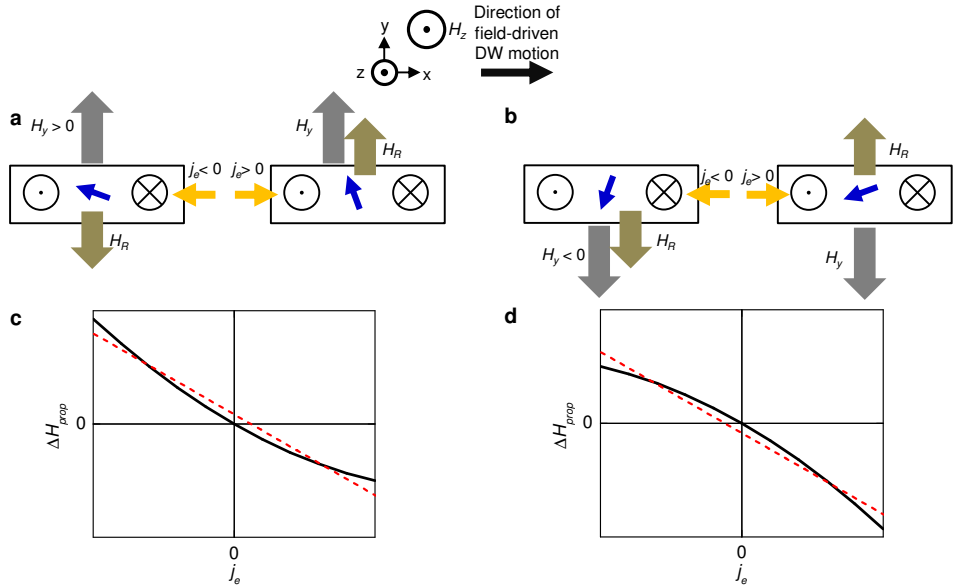

**Figure S7**. (a,b) Illustrations of the effect of the Rashba field $H_R$ on the DW magnetization for (a) $H_y > 0$ and (b) $H_y < 0$. (c,d) Illustrations of the effect $H_R$ would have on measurements of $\Delta H_{prop}/j_e$ (solid black curve) for (c) $H_y > 0$ and (d) $H_y < 0$. Note that even if a Rashba field were present, it would not account for the asymmetry in Fig. S8 since $H_y > 0$ and $H_y < 0$ would exhibit the same slope if a linear fit were used to fit in $\Delta H_{prop}$ against $j_e$ (dotted red line). Note also that the curvature in these schematically represented data is not present in the actual experimental data, such as are shown in Fig. 2(d)-(f) of the main text.

The current-induced field from the Rashba effect, or any other effective transverse field that scales linearly with $j_e$, cannot account for the asymmetry in the efficiency $\chi$ versus $H_y$, for the following reasons:

1) The nonlinear relation between $\Delta H_{prop}$ and $j_e$, expected under a strong Rashba field, is not observed in the experimental data (Fig. 2(d)-(f) in the main text, or in Figs. S3 and S4 above.

2) Even if this nonlinear relation (i.e., Rashba field) were present, $\chi$ would be identical for both polarities of $H_y$ (Fig. S7).

In these experiments, an out-of-plane driving field $H_z$ is applied, which acts to depin the DW and drive it along the nanostrip, with the SHE effective field either assisting or impeding the field-driven motion. Although the experiment is close to the quasistatic regime due to the low current densities and long timescales, the propagation field nonetheless exerts a torque on the DW moment as the DW moves. This torque is proportional to $\hat{m} \times H_z \hat{z}$. In the experiments, the sign of $H_z$ is reversed to drive up-down and down-up DWs in the same direction along the nanostrip for detection by MOKE. Since $\hat{m}$ changes sign also for up-down and down-up DWs due to the chirality, the field torque tends to cant the DW moment in the same direction along the y-axis for up-down and down-up DWs. In the depinning field measurements, when $H_y$ is aligned with the $H_z$-induced projection of $\hat{m}$ along y, the depinning efficiency is more easily reduced than when $H_y$ is oriented in the opposite direction (see Figs. S7 and S8).

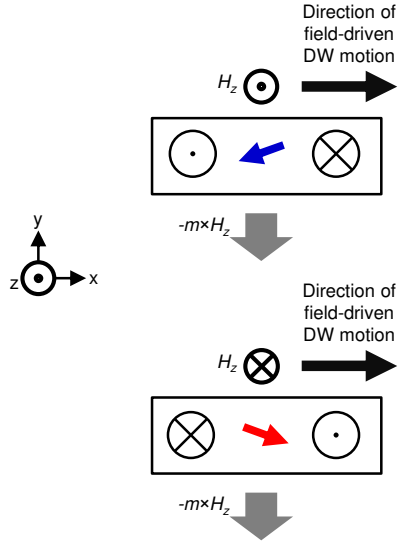

**Figure S8**. Slight rotation of the internal DW moment due to the $H_z$-induced torque. Both the up-down and down-up DWs move quasistatically in the +$x$-direction.

**References**
[1] S. Emori, U. Bauer, S.-M. Ahn, E. Martinez, and G. S. D. Beach, Nat. Mater. **12**, 611 (2013).
[2] S. Emori, D. C. Bono, and G. S. D. Beach, J. Appl. Phys. **111**, 07D304 (2012).
[3] S. Lemerle, J. Ferré, C. Chappert, V. Mathet, T. Giamarchi, and P. Le Doussal, Phys. Rev. Lett. **80**, 849 (1998).
[4] P. J. Metaxas, J. P. Jamet, A. Mougin, M. Cormier, J. Ferré, V. Baltz, B. Rodmacq, B. Dieny, and R. L. Stamps, Phys. Rev. Lett. **99**, 217208 (2007).
[5] E. Martinez, J. Phys. Condens. Matter **24**, 024206 (2012).
[6] S.-G. Je, D.-H. Kim, S.-C. Yoo, B.-C. Min, K.-J. Lee, and S.-B. Choe, Phys. Rev. B **88**, 214401 (2013).
[7] T. Suzuki, S. Fukami, N. Ishiwata, M. Yamanouchi, S. Ikeda, N. Kasai, and H. Ohno, Appl. Phys. Lett. **98**, 142505 (2011).
[8] J. Kim, J. Sinha, M. Hayashi, M. Yamanouchi, S. Fukami, T. Suzuki, S. Mitani, and H. Ohno, Nat. Mater. **12**, 240 (2013).